\documentstyle[prl,aps]{revtex}
\newcommand{\be}{\begin{equation}}
\newcommand{\ee}{\end{equation}}
\newcommand{\ba}{\begin{eqnarray}}
\newcommand{\ea}{\end{eqnarray}}

\def\a{\alpha}
\def\b{\beta}

\def\d{\delta}

\def\ve{\varepsilon}
\def\f{\phi}
\def\vf{\varphi}

\def\h{\eta}

\def\l{\lambda}
\def\m{\mu}
\def\n{\nu}
\def\p{\pi}

\def\r{\rho}

\def\t{\tau}

\def\D{\Delta}
\def\F{\Phi}
\def\G{\Gamma}

\def\P{\Pi}
\def\Q{\Theta}
\def\S{\Sigma}

\def\ca{{\cal A}}

\def\cf{{\cal F}}
\def\cg{{\cal G}}

\def\ck{{\cal K}}

\def\co{{\cal O}}
\def\cp{{\cal P}}

\def\cs{{\cal S}}

\def\cw{{\cal W}}

\newcommand{\ov}{\overline}
\newcommand{\ti}{\tilde}

\newcommand{\wt}{\widetilde}
\newcommand{\wh}{\widehat}

\newcommand{\aand}{\;\;\;\mbox{and}\;\;\;}
\newcommand{\pa}{\partial}

\def\bc{\bar c}

\begin{document}
\draft
\title{On the finiteness of a new topological model in $D=3$}
\author{O.M. Del Cima$^a$\thanks{Supported by the 
{\it Fonds zur 
F\"orderung der Wissenschaftlichen Forschung (FWF)} under the contract 
number P11654-PHY. E-mail:{\tt delcima@tph73.tuwien.ac.at}.}, 
J.M. Grimstrup$^{a,b}$\thanks{E-mail:{\tt jesper@tph41.tuwien.ac.at}.} and 
M. Schweda$^a$\thanks{E-mail:{\tt mschweda@tph.tuwien.ac.at}.}   }
\address{
$^a${\it Institut f\"ur Theoretische Physik (ITP),}\\
Technische Universit\"at Wien (TU-Wien),\\
Wiedner Hauptstra{\ss}e 8-10 - A-1040 - Vienna - Austria.\\  
$^b${\it The Niels Bohr Institute (NBI),}\\
Blegdamsvej 17 - DK-2100 - Copenhagen {\O} - Denmark. } 
\date{\today}
\maketitle

\begin{abstract}

A new topological model is proposed in three dimensions as an extension of 
the BF-model. It is a three-dimensional counterpart of the two-dimensional 
model introduced by Chamseddine and Wyler ten years ago. The BFK-model, 
as we shall call it, shows to be quantum scale invariant at all orders in 
perturbation theory. The proof of its full finiteness is given in the 
framework of algebraic renormalization.
 
\end{abstract}

\pacs{PACS numbers: 11.10.Gh 11.10.Kk 11.15.-q 11.15.Bt  
\hspace{4,90cm}TUW-99-12}

\section{Introduction.}

The BF-models have been widely investigated since the middle of the eighties 
\cite{birmingham,fukuyama,chamseddine,blau,blasi,piguet1,schweda,piguet2,stephany,henneaux,accardi,delcima}, specially the 
two-dimensional BF-model in connection with topological quantum gravity 
\cite{birmingham,fukuyama,chamseddine,blau}. 
The properties of ultraviolet and infrared perturbative finiteness of 
the two-dimensional BF-model have been rigorously proved in the framework 
of algebraic renormalization \cite{blasi}, there the authors make use of 
an additional symmetry peculiar to topological models in the Landau gauge, 
called vector-supersymmetry \cite{piguet1,delduc,angra,sorella}. As 
a natural extension of the two-dimensional case, it has been proposed in 
\cite{chamseddine} an extended BF-model, as a gauge theory for topological 
quantum gravity, which accommodates a topological matter coupling. Later, 
the authors of \cite{schweda} have been shown that the model proposed in 
\cite{chamseddine} is still ultraviolet and infrared finite at all orders 
in perturbation theory by using the same approach as those of \cite{blasi} 
have done to the pure two-dimensional BF-model.

At the level of perturbation theory, a general classification of all 
possible anomalies and invariant counterterms of BF-models in any space-time 
dimension has been done in \cite{piguet2}. The feature of the independence 
on the gauge coupling of those models was pointed out by the authors of 
\cite{stephany}. Moreover, it has been shown that $D$-dimensional 
BF-Yang-Mills-models, a BF formulation of the Yang-Mills theory, are 
cohomologically equivalent to Yang-Mills \cite{henneaux}. Three-dimensional 
BF-Yang-Mills theory, both Gaussian and extended, have been algebraically 
quantized by the authors of \cite{accardi}, where the proof of their exact 
quantum scale invariance at all orders in perturbation theory have been 
given in \cite{delcima}. Note: after the completion of this work we 
became aware of ref.\cite{renan}, where a dimensional reduction 
{\it \`a la} Scherk of the Gaussian four-dimensional Abelian version 
of the BF-Yang-Mills-model is performed to three dimensions leading to 
a model similar to ours, there its spectrum is analyzed in details as well.    

Our purpose in this letter is to present a rigorous proof on the 
finiteness of an extended version of the BF-model in three dimensions 
introduced here. Such a model, we shall call BFK-model, is a three-dimensional 
counterpart of that model in two dimensions proposed by 
Chamseddine and Wyler \cite{chamseddine}. 
The proof is performed by using the method of algebraic renormalization 
\cite{piguet1}, which is independent of any particular regularization scheme.
The letter is organized as follows. The BFK-model and its symmetries are 
introduced in Section II. In Section III, the proof on the full finiteness 
at all orders in perturbation theory is sketched and at the end the 
conclusions are drawn. Here in this letter we summarize the main 
results and all details shall be reported elsewhere in a more complete 
paper \cite{paper}. It is now under investigation the classification 
of all possible counterterms and anomalies of the $D$-dimensional BFK-model 
\cite{d-bfk} by following the same approach of \cite{piguet2}. 

\section{The BFK-model in $D=3$ and its symmetries}
\subsection{The classical action}
The classical action of the BFK-model in $D=3$ is given by
\be
\S_{\rm{inv}}^{\rm{BFK}}=\frac 12~{\mbox{Tr}}\int d^3x~
\ve^{\m\n\r}\left\{ B_{\m}F_{\n\r} + K_{\m\n}D_\r\f \right\}~, 
\label{bfkinv}
\ee
where $B_{\m}$ is a vector field, $K_{\m\n}$ is a rank-2 antisymmetric 
tensor and $\f$ is a scalar. 
The second piece of the action (\ref{bfkinv}) can be seen as a topological 
matter coupling, where the matter fields $K_{\m\n}$ and $\f$ lie in the 
adjoint representation of the gauge group. In two dimensions similar 
topological matter term was proposed by Chamseddine and Wyler 
\cite{chamseddine}. Any field, $\vf$, is to be assumed as Lie algebra 
valued, in such a way that  
\be
\vf\equiv\vf^a\t_a~, 
\ee
where the matrices $\t$ are the generators of the gauge group\footnote{The 
gauge group is considered as a general compact one.} and obey
\be
[\t_a,\t_b]=f_{abc}\t_c \aand {\mbox{Tr}}(\t_a\t_b)=\frac 12 \d_{ab}~.
\ee

The field strength, $F_{\m\n}$, is defined as  
\be
F_{\m\n}=\pa_\m A_\n-\pa_\n A_\m + [A_\m,A_\n]~,
\label{fs}
\ee
and the covariant derivative reads
\be
D_\m\f=\pa_\m \f + [A_\m,\f]~.
\ee

\subsection{Gauge symmetries}
The action (\ref{bfkinv}) possesses two symmetries:
\begin{enumerate}
\item The standard gauge symmetry
\ba
&&\d_\a A_\m=-D_\m \a \equiv -(\pa_\m \a + [A_\m,\a])~,~~
\d_\a B_\m=[\a,B_\m]~,\nonumber \\
&&\d_\a K_{\m\n}=[\a,K_{\m\n}]\aand\d_\a \f=[\a,\f]~.\label{gsym}
\ea
\item The topological symmetry
\ba
&&\d_\b A_\m=0~,~~\d_\b B_\m=-(D_\m \b + [\b_\m,\f])~,\nonumber \\
&&\d_\b K_{\m\n}=-(D_\m \b_\n - D_\n \b_\m)\aand\d_\b \f=0~.
\label{tsym}
\ea
\end{enumerate}

\subsection{BRS symmetry}
The corresponding BRS transformations of the fields, $A_\m$, $B_\m$, 
$K_{\m\n}$ and $\f$, stemming from the symmetries (\ref{gsym}) and 
(\ref{tsym}), are given by\footnote{The commutators are assumed to 
be graded, namely, $[\vf_1^{g_1},\vf_2^{g_2}]
\equiv\vf_1^{g_1}\vf_2^{g_2}-(-1)^{g_1.g_2}\vf_2^{g_2}\vf_1^{g_1}$, where the 
upper indices, $g_1$ and $g_2$, are the Faddeev-Popov charges ($\F\P$) carried 
by $\vf_1^{g_1}$ and $\vf_2^{g_2}$, respectively.}
\ba
&&sA_\m=-D_\m c \equiv -(\pa_\m c + [A_\m,c])~,~~
sB_\m=-D_\m B^1 + [\f,B^1_\m] + [c,B_\m]~,\nonumber \\
&&sK_{\m\n}=-(D_\m B^1_\n - D_\n B^1_\m) + [c,K_{\m\n}]~,~~
s\f=[c,\f]\aand sc=c^2~,
\label{BRS0}
\ea
where $c$ and $B^1$ are scalar ghosts, and $B^1_\m$ is a vector 
ghost, all of them are anticommuting fields with Faddeev-Popov charge 
(ghost number) one. Bearing in mind that we have a residual degree 
of freedom from $sK_{\m\n}$ caused by a zero mode ($B^1_\m=D_\m\vf^2$), 
{\it i.e.}, it is a reducible symmetry, has to be fixed, therefore, yielding 
a ghost $B^2$ for the ghost $B^1_\m$. Now, fixing the zero mode 
by introducing the ghost for ghost, $B^2$, the remaining BRS 
transformations read
\be
sB^1=[\f,B^2] + [c,B^1]~,~~sB^1_\m=D_\m B^2 + [c,B^1_\m]\aand 
sB^2=[c,B^2]~. 
\label{BRS1}
\ee

It should be noticed that the BRS operator $s$ is nilpotent up to 
the field equations for $B_\m$ and $K_{\m\n}$, since
\be
s^2B_\m=
\frac12 \left[B^2,
\ve_{\m\n\r}{\frac{\d \S_{\rm{inv}}^{\rm{BFK}}}{\d K_{\n\r}}}\right]
\aand
s^2K_{\m\n}=
\left[B^2,
\ve_{\m\n\r}{\frac{\d \S_{\rm{inv}}^{\rm{BFK}}}{\d B_\r}}\right]~,
\label{onshell}
\ee 
called on-shell nilpotency.

\subsection{Gauge-fixing}
The gauge-fixing we are considering here is of the Landau-type. Since we 
are dealing with a more complex model than the Yang-Mills one, face 
its symmetries, some subtleties arise as their consequence. In order to 
implement the gauge-fixing we couple the Lagrange multiplier fields $b$, 
$\p^0$, $\p^{0\m}$ and $\p^{-1}$ to 
\be
{\mbox{Tr}}~b\pa^\m A_\m~,~~{\mbox{Tr}}~\p^0\pa^\m B_\m~,~~
{\mbox{Tr}}~\p^{0\n}(\pa^\m K_{\m\n} + \pa_\n\r^0)\aand
{\mbox{Tr}}~\p^{-1}(\pa^\m B^1_\m + \l^1)~,
\label{gaugecond}
\ee
where the multiplier fields belong to the following BRS-doublets:
\ba
s\bc&=&b~,~~sb=0~; \nonumber \\
s\bc^{-1}&=&\p^0~,~~s\p^0=0~; \nonumber \\
s\bc^{-1\m}&=&\p^{0\m}~,~~s\p^{0\m}=0~; \nonumber \\
s\bc^{-2}&=&\p^{-1}~,~~s\p^{-1}=0~.
\label{doublets}
\ea
We stress here that for the fields, $K_{\m\n}$ and $B^1_\m$, 
inhomogeneous gauge conditions (\ref{gaugecond}) have been chosen. In which 
concerns the field $K_{\m\n}$, a gauge condition of the type 
$\p^{0\n}\pa^\m K_{\m\n}$ would not fix completely the gauge, since a 
residual degree of freedom is present by $\p^{0\n}\rightarrow\p^{0\n}
+\pa^\n\ti\r^0$, therefore, due to this fact an inhomogeneous 
gauge-fixing condition has to be adopted. Since it is an Abelian 
transformation does not demand further ghost fields. Besides the condition 
on $K_{\m\n}$ has introduced a new field $\r^0$, it yields also 
the consideration of another inhomogeneous gauge condition associated to 
$B^1_\m$ by putting into the game the field $\l^1$. In fact, the 
introduction of $\l^1$ is due to cohomological arguments, so in order to 
protect the independence of BRS-cohomology in those fields introduced by hand, 
we force them to belong to a BRS-doublet
\be
s\r^0=~\l^1~,~~s\l^1=0~.
\label{doublet}
\ee 
Bearing in mind that a neutral Faddeev-Popov charge action is desired, the 
fields $\r^0$ and $\l^1$ have already had their charges 
fixed\footnote{The dimension (d) and the ghost number ($\F\P$) of all fields 
are displayed in TABLE \ref{dimensions}.}, $0$ and $1$, 
respectively.

Now, to introduce a BRS-trivial gauge-fixing compatible with the gauge 
conditions (\ref{gaugecond}), we add to the action (\ref{bfkinv}) the 
following four pieces:
\ba
\S_{\rm{gf}}^{\rm{1}}&=&s~{\mbox{Tr}}\int d^3x~\bc\pa^\m A_\m~,~~
\S_{\rm{gf}}^{\rm{2}}=s~{\mbox{Tr}}\int d^3x~\bc^{-1}
\pa^\m B_\m~,\nonumber \\
\S_{\rm{gf}}^{\rm{3}}&=&s~{\mbox{Tr}}\int d^3x~\bc^{-1\n}
(\pa^\m K_{\m\n} + \pa_\n\r^0)\aand
\S_{\rm{gf}}^{\rm{4}}=s~{\mbox{Tr}}\int d^3x~\bc^{-2}
(\pa^\m B^1_\m + \l^1)~.
\label{gaugefix}
\ea
Moreover, since those four pieces (\ref{gaugefix}) of the gauge-fixing 
shall break the on-shell nilpotency (\ref{onshell}) of the BRS operator $s$, 
further modifications of the BRS transformations are necessary, however, 
in order to restore the BRS invariance an additional term has to be added to 
the gauge-fixing sector as well. The full modified gauge-fixing action, 
\be
\S_{\rm{gf}}=\S_{\rm{gf}}^{\rm{1}}+
\S_{\rm{gf}}^{\rm{2}}+\S_{\rm{gf}}^{\rm{3}}+
\S_{\rm{gf}}^{\rm{4}}+\S_{\rm{mod}}~, 
\ee
with
\be
\S_{\rm{mod}}=
{\mbox{Tr}}\int d^3x~\ve^{\m\n\r}\pa_\m\bc^{-1}_\n[\pa_\r\bc^{-1},B^2]~,
\ee
reads
\ba
\S_{\rm{gf}}&=&
{\mbox{Tr}}\int d^3x~\biggl\{b\pa^\m A_\m + \p^0\pa^\m B_\m +
\p^{0\n}(\pa^\m K_{\m\n} + \pa_\n\r^0) + \p^{-1}(\pa^\m B^1_\m + \l^1) + 
\nonumber \\
&+&\bc\pa^\m D_\m c + 
\bc^{-1}\pa^\m(D_\m B^1 - [\f,B^1_\m] - [c,B_\m] - 
\ve_{\m\n\r}[\pa^\n\bc^{-1\r},B^2]) +
\nonumber \\
&+&\bc^{-1\n}[\pa^\m(D_\m B^1_\n - D_\n B^1_\m 
- [c,K_{\m\n}] - \ve_{\m\n\r}[\pa^\r\bc^{-1},B^2]) - \pa_\n\l^1] +
\nonumber \\
&+&\bc^{-2}\pa^\m(D_\m B^2 + [c,B^1_\m]) + 
\ve^{\m\n\r}\pa_\m\bc^{-1}_\n[\pa_\r\bc^{-1},B^2]\biggr\}~, 
\label{bfkgf}
\ea
where the enlarged BRS transformations associated to $B_\m$ and $K_{\m\n}$ 
are given by
\ba
sB_\m&=&-D_\m B^1 + [\f,B^1_\m] + [c,B_\m] + 
\ve_{\m\n\r}[\pa^\n\bc^{-1\r},B^2]~,  \nonumber \\
sK_{\m\n}&=&-(D_\m B^1_\n - D_\n B^1_\m) + [c,K_{\m\n}] + 
\ve_{\m\n\r}[\pa^\r\bc^{-1},B^2]~. 
\ea

Let us now introduce the action in which the nonlinear BRS transformations 
are coupled to the antifields (BRS invariant external fields), so as to 
control the renormalization of those transformations:
\ba
\S_{\rm{ext}}&=&{\mbox{Tr}}\int d^3x~\biggl\{B^*_\m sB^\m + 
\frac12 K^*_{\m\n}sK^{\m\n} + A^*_\m sA^\m + \f^*s\f + c^*sc + B^{2*}sB^2 
+ B^{1*}sB^1 + B^{1*}_\m sB^{1\m} + \nonumber \\
&+&{\frac 12}\ve^{\m\n\r}B^*_\m[K^*_{\n\r},B^2]\biggr\}~.\label{bfkext}
\ea

The total classical action for the BFK-model, $\G^{(0)}$:
\be
\G^{(0)}=\S_{\rm{inv}}^{\rm{BFK}}+\S_{\rm{gf}}+\S_{\rm{ext}}~,
\label{bfkaction} 
\ee
is invariant under the following BRS transformations
\ba
&sB_\m=-D_\m B^1 + [\f,B^1_\m] + [c,B_\m] + 
\ve_{\m\n\r}[\pa^\n\bc^{-1\r},B^2]~,  \nonumber \\
&sK_{\m\n}=-(D_\m B^1_\n - D_\n B^1_\m) + [c,K_{\m\n}] + 
\ve_{\m\n\r}[\pa^\r\bc^{-1},B^2]~, \nonumber \\ 
&sA_\m=-D_\m c~,~~s\f=[c,\f]~,~~sc=c^2~, \nonumber \\
&sB^1=[\f,B^2] + [c,B^1]~,~~sB^1_\m=D_\m B^2 + [c,B^1_\m]~,~~
sB^2=[c,B^2]~,\nonumber \\
&s\bc=b~,~~sb=0~;~~s\bc^{-1}=\p^0~,~~s\p^0=0~; \nonumber \\
&s\bc^{-1\m}=\p^{0\m}~,~~s\p^{0\m}=0~;~~
s\bc^{-2}=\p^{-1}~,~~s\p^{-1}=0~,\nonumber \\
&s\r^0=\l^1~,~~s\l^1=0~.
\label{BRStotal}
\ea

The BRS invariance of the action (\ref{bfkaction}) is expressed through 
the Slavnov-Taylor identity
\ba
\cs(\G^{(0)})&=&{\mbox{Tr}}\int d^3x~\biggl\{
{\frac{\d\G^{(0)}}{\d B^*_\m}}{\frac{\d\G^{(0)}}{\d B^\m}}  +
\frac12 {\frac{\d\G^{(0)}}{\d K^*_{\m\n}}}{\frac{\d\G^{(0)}}{\d K^{\m\n}}} +
{\frac{\d\G^{(0)}}{\d A^*_\m}}{\frac{\d\G^{(0)}}{\d A^\m}} +
{\frac{\d\G^{(0)}}{\d \f^*}}{\frac{\d\G^{(0)}}{\d \f}} + 
{\frac{\d\G^{(0)}}{\d c^*}}{\frac{\d\G^{(0)}}{\d c}} +
{\frac{\d\G^{(0)}}{\d B^{2*}}}{\frac{\d\G^{(0)}}{\d B^2}} + \nonumber \\
&+& {\frac{\d\G^{(0)}}{\d B^{1*}}}{\frac{\d\G^{(0)}}{\d B^1}} + 
{\frac{\d\G^{(0)}}{\d B^{1*}_\m}}{\frac{\d\G^{(0)}}{\d B^{1\m}}} +
b{\frac{\d\G^{(0)}}{\d \bc}} + \p^0{\frac{\d\G^{(0)}}{\d\bc^{-1}}} +
\p^{0\m}{\frac{\d\G^{(0)}}{\d\bc^{-1\m}}} + 
\p^{-1}{\frac{\d\G^{(0)}}{\d\bc^{-2}}} + 
\l^1{\frac{\d\G^{(0)}}{\d \r^0}}\biggr\}= 0~,\label{slavnov}
\ea
which translates, in a functional way, the invariance of the classical theory 
under the BRS symmetry. It is suitable to define, for later use, 
the linearized Slavnov-Taylor operator as below
\ba
\cs_{\G^{(0)}}&=&{\mbox{Tr}}\int d^3x~\biggl\{
{\frac{\d\G^{(0)}}{\d B^*_\m}}{\frac{\d}{\d B^\m}} +
{\frac{\d\G^{(0)}}{\d B^\m}}{\frac{\d}{\d B^*_\m}} +
\frac12 {\frac{\d\G^{(0)}}{\d K^*_{\m\n}}}{\frac{\d}{\d K^{\m\n}}} +
\frac12 {\frac{\d\G^{(0)}}{\d K^{\m\n}}}{\frac{\d}{\d K^*_{\m\n}}} +
{\frac{\d\G^{(0)}}{\d A^*_\m}}{\frac{\d}{\d A^\m}} +
{\frac{\d\G^{(0)}}{\d A^\m}}{\frac{\d}{\d A^*_\m}} + \nonumber \\
&+&{\frac{\d\G^{(0)}}{\d \f^*}}{\frac{\d}{\d \f}} + 
{\frac{\d\G^{(0)}}{\d \f}}{\frac{\d}{\d \f^*}} +
{\frac{\d\G^{(0)}}{\d c^*}}{\frac{\d}{\d c}} +
{\frac{\d\G^{(0)}}{\d c}}{\frac{\d}{\d c^*}} +
{\frac{\d\G^{(0)}}{\d B^{2*}}}{\frac{\d}{\d B^2}} + 
{\frac{\d\G^{(0)}}{\d B^2}}{\frac{\d}{\d B^{2*}}} +
{\frac{\d\G^{(0)}}{\d B^{1*}}}{\frac{\d}{\d B^1}} +
{\frac{\d\G^{(0)}}{\d B^1}}{\frac{\d}{\d B^{1*}}} +
\nonumber \\
&+& {\frac{\d\G^{(0)}}{\d B^{1*}_\m}}{\frac{\d}{\d B^{1\m}}} +
{\frac{\d\G^{(0)}}{\d B^{1\m}}}{\frac{\d}{\d B^{1*}_\m}} +
b{\frac{\d}{\d \bc}} + \p^0{\frac{\d}{\d\bc^{-1}}} +
\p^{0\m}{\frac{\d}{\d\bc^{-1\m}}} + \p^{-1}{\frac{\d}{\d\bc^{-2}}} + 
\l^1{\frac{\d}{\d \r^0}}\biggr\}~.\label{slavnovlin}
\ea

Since the BFK-model is a topological model of Schwarz type, in the 
Landau gauge, as we are assuming here, a symmetry called vector-supersymmetry 
\cite{piguet1,delduc,angra,sorella} stems:
\be
\cw_\m \G^{(0)}=\D^{\cw_\m}_\m~,
\ee 
where its representation by means of a Ward operator is given by
\ba
\cw_\m&=&{\mbox{Tr}}\int d^3x~\biggl\{ 
-\ve_{\m\n\r}(\pa^\n\bc+A^{*\n})\frac{\d}{\d B_\r}
-\frac12 \ve_{\m\n\r}\f^*\frac{\d}{\d K_{\n\r}}
-\ve_{\m\n\r}(\pa^\n\bc^{-1}+B^{*\n})\frac{\d}{\d A_\r} 
+\frac12\ve_{\m\n\r}(\pa^{[\n}\bc^{-1\r]}+K^{*\n\r})\frac{\d}{\d\f}+ 
\nonumber\\
&-&A_\m\frac{\d}{\d c}+B^1_\m\frac{\d}{\d B^2}-B_\m\frac{\d}{\d B^1}
-K_{\m\n}\frac{\d}{\d B^1_{\n}}-\bc^{-2}\frac{\d}{\d \bc^{-1\m}}
+\pa_\m\bc\frac{\d}{\d b}+\pa_\m\bc^{-1}\frac{\d}{\d \p^0}
+(\pa_\m\bc^{-1}_\n+\h_{\m\n}\p^{-1})\frac{\d}{\d \p^0_{\n}}+
\nonumber\\
&+&\pa_\m\bc^{-2}\frac{\d}{\d \p^{-1}}+\pa_\m\r^0\frac{\d}{\d \l^1}
-B^{1*}\frac{\d}{\d B^{*\m}}
-\frac12 \h_{\m[\n}B^{1*}_{\r]}\frac{\d}{\d K^*_{\n\r}}
-c^*\frac{\d}{\d A^{*\m}}-B^{2*}\frac{\d}{\d B^{1*\m}}\biggr\}~,
\ea
and the classical breaking (linear in the quantum fields), $\D^{\cw_\m}_\m$, 
reads
\ba
\D^{\cw_\m}_\m&=&{\mbox{Tr}}\int d^3x~\biggl\{
-B^*_\n\pa_\m B^\n-\frac12 K^*_{\n\r}\pa_\m K^{\n\r}-A^*_\n\pa_\m A^\n
-\f^*\pa_\m\f+c^*\pa_\m c-B^{2*}\pa_\m B^2+B^{1*}\pa_\m B^1
+B^{1*}_\n\pa_\m B^{1\n}+
\nonumber\\
&-&\ve_{\m\n\r}B^{*\n}\pa^\r b+\ve_{\m\n\r}A^{*\n}\pa^\r \p^0
+\ve_{\m\n\r}\f^*\pa^\n \p^{0\r}\biggr\}~.
\ea
By similar reasons to the existence of a vector-supersymmetry in the 
BFK-model, a scalar-supersymmetry can be found:
\be
\L \G^{(0)}=\D^{\L}~,
\ee
where its Ward operator reads
\ba
\L&=&{\mbox{Tr}}\int d^3x~\biggl\{
\frac12 \ve_{\m\n\r}(\pa^\m\bc+A^{*\m})\frac{\d}{\d K_{\n\r}} 
-\frac12\ve_{\m\n\r}(\pa^{[\n}\bc^{-1\r]}+K^{*\n\r})\frac{\d}{\d A_\m}
-\f\frac{\d}{\d c}+B^1\frac{\d}{\d B^2}+B_\m\frac{\d}{\d B^1_{\m}}+
\nonumber\\
&-&\bc^{-2}\frac{\d}{\d \bc^{-1}}+\p^{-1}\frac{\d}{\d \p^0}
+B^{1*}_\m\frac{\d}{\d B^*_\m}-c^*\frac{\d}{\d\f^*}
-B^{2*}\frac{\d}{\d B^{1*}}\biggr\}~,
\ea
and its linear breaking in the quantum fields, $\D^{\L}$, is given by
\be
\D^{\L}={\mbox{Tr}}\int d^3x~\biggl\{\frac12\ve_{\m\n\r}\left(
K^{*\m\n}\pa^\r b-A^{*\m}\pa^{[\n}\p^{0\r]}\right)\biggr\}~.
\ee

\subsection{Gauge conditions, ghost and antighost equations}
This Subsection is devoted to establish the gauge conditions, 
ghost and antighost equations, and two others quite important 
symmetries in the proof of the exact quantum scale invariance of the BFK-model.

The gauge conditions read
\be
\frac{\d\G^{(0)}}{\d b}=\pa^\m A_\m~,~~
\frac{\d\G^{(0)}}{\d \p^0}=\pa^\m B_\m~,~~
\frac{\d\G^{(0)}}{\d \p^{0\n}}=\pa^\m K_{\m\n}+\pa_\n\r^0\aand
\frac{\d\G^{(0)}}{\d \p^{-1}}=\pa^\m B^1_\m+\l^1~,
\ee
moreover, the conditions fulfilled by the BRS-doublet auxiliary fields, 
$\r^0$ and $\l^1$, are given by
\be
\frac{\d\G^{(0)}}{\d \r^0}=-\pa^\m \p^0_\m\aand
\frac{\d\G^{(0)}}{\d \l^1}=-\pa^\m \bc^{-1}_\m-\p^{-1}~.
\ee
The ghost equations: 
\ba
&&\cg(\G^{(0)})\equiv\frac{\d\G^{(0)}}{\d \bc}
+\pa^\m\frac{\d\G^{(0)}}{\d A^{*\m}}=0~,~~
\cg^1(\G^{(0)})\equiv\frac{\d\G^{(0)}}{\d \bc^{-1}}
+\pa^\m\frac{\d\G^{(0)}}{\d B^{*\m}}=0~,\nonumber\\
&&\cg^1_\m(\G^{(0)})\equiv\frac{\d\G^{(0)}}{\d \bc^{-1\m}}
+\pa^\n\frac{\d\G^{(0)}}{\d K^{*\n\m}}=-\pa_\m\l^1\aand
\cg^2(\G^{(0)})\equiv\frac{\d\G^{(0)}}{\d \bc^{-2}}
-\pa^\m\frac{\d\G^{(0)}}{\d B^{1*\m}}=0~,
\ea
mean that $\G^{(0)}$ depends on the antighosts, $\bc$, $\bc^{-1}$, 
$\bc^{-1}_\m$ and $\bc^{-2}$, and the antifields, $A^*_\m$, $B^*_\m$, 
$K^*_{\m\n}$ and $B^{1*}_\m$, through the combinations
\be
{\wt A}^*_\m=A^*_\m+\pa_\m\bc~,~~{\wt B}^*_\m=B^*_\m+\pa_\m\bc^{-1}~,~~
{\wt K}^*_{\m\n}=K^*_{\m\n}+\pa_{[\m}\bc^{-1}_{\n]}\aand
{\wt B}^{1*}_\m=B^{1*}_\m-\pa_\m\bc^{-2}~.
\ee

In the BFK-model there are two antighost equations, they are listed 
as below:
\ba
\ov\cg(\G^{(0)})&\equiv&\int d^3x~\biggl\{\frac{\d\G^{(0)}}{\d B^1}
+\biggl[\bc^{-1},\frac{\d\G^{(0)}}{\d b}\biggr]\biggr\}=\D^{\ov\cg}~,~~
{\mbox{where}}\nonumber\\
\D^{\ov\cg}&\equiv&\int d^3x~\{[A^\m,B^*_\m]+[c,B^{1*}]\}~,\\
\AE(\G^{(0)})&\equiv&\int d^3x~\biggl\{\frac{\d\G^{(0)}}{\d B^2}
-\biggl[\bc^{-2},\frac{\d\G^{(0)}}{\d b}\biggr]\biggr\}
=\D^{\AE}~,~~{\mbox{where}}\nonumber\\
\D^{\AE}&\equiv&\int d^3x~\biggl\{\frac12\ve_{\m\n\r}
\left([B^{*\m},(\pa^{[\n}\bc^{-1\r]}+K^{*\n\r})]
+[K^{*\m\n},\pa^\r\bc^{-1}]\right)+[B^{2*},c]+[B^{1*},\f]
+[B^{1*}_\m,A^\m]\biggr\}~.
\ea
From those operators, $\ov\cg$ and $\AE$, two others can be found by grading 
commutations with the Slavnov-Taylor operator (see the operatorial 
algebra in the next Subsection): 
\ba
\ov\cf(\G^{(0)})&\equiv&\int d^3x~\biggl\{ 
\biggl[A_\m,\frac{\d\G^{(0)}}{\d B_\m}\biggr]
-\biggl[B^*_\m,\frac{\d\G^{(0)}}{\d A^*_\m}\biggr]
+\biggl[c,\frac{\d\G^{(0)}}{\d B^1}\biggr]
+\biggl[B^{1*},\frac{\d\G^{(0)}}{\d c^*}\biggr]
+\biggl[\bc^{-1},\frac{\d\G^{(0)}}{\d \bc}\biggr]+\nonumber\\
&+&\biggl[\p^0,\frac{\d\G^{(0)}}{\d b}\biggr]\biggr\}=0~, \\
\O(\G^{(0)})&\equiv&\int d^3x~\biggl\{ 
-\frac12\ve_{\m\n\r}
\left(\biggl[(\pa^{[\m}\bc^{-1\n]}+K^{*\m\n}),
\frac{\d\G^{(0)}}{\d B_\r}\biggr]
+\biggl[(\pa^\m\bc^{-1}+B^{*\m}),
\frac{\d\G^{(0)}}{\d K_{\n\r}}\biggr]\right)
-\biggl[c,\frac{\d\G^{(0)}}{\d B^2}\biggr]
-\biggl[\f,\frac{\d\G^{(0)}}{\d B^1}\biggr]+ \nonumber\\ 
&-&\biggl[A_\m,\frac{\d\G^{(0)}}{\d B^1_\m}\biggr]
-\biggl[\bc^{-2},\frac{\d\G^{(0)}}{\d \bc}\biggr]
+\biggl[\p^{-1},\frac{\d\G^{(0)}}{\d b}\biggr]
+\biggl[B^{1*}_\m,\frac{\d\G^{(0)}}{\d A^*_\m}\biggr]
+\biggl[B^{1*},\frac{\d\G^{(0)}}{\d \f^*}\biggr]
-\biggl[B^{2*},\frac{\d\G^{(0)}}{\d c^*}\biggr]
\biggr\}=\D^{\O}\nonumber\\
{\mbox{where}}~~\D^{\O}&\equiv&\int d^3x~\biggl\{ 
\frac12\ve_{\m\n\r}\left([\p^0,\pa^\m K^{*\n\r}]
- [\p^{0\m},\pa^{[\n} B^{*\r]}]\right)\biggr\}~.
\ea
It should be pointed out that the breakings, $\D^{\ov\cg}$, $\D^{\AE}$ 
and $\D^{\O}$, being linear in the quantum fields are not subjected to 
renormalization.

\subsection{Operatorial algebra}
All operators introduced previously satisfy the following off-shell 
algebra for any functional $\ck$ with even Faddeev-Popov charge:
\begin{enumerate}
\item Slavnov-Taylor operator identities
\ba
&&S_{\ck}S(\ck)=0~\forall\ck~,~~S_{\ck}S_{\ck}=0~{\rm if}~S(\ck)=0~,\nonumber\\
&&\cw_{\m}S(\ck)
+S_{\ck}(\cw_{\m}(\ck)-\D^{\cw_\m}_{\m})=\cp_{\m}(\ck)~,~~{\L}S(\ck)
+S_{\ck}({\L}(\ck)-\D^{\L})=0~,\nonumber\\
&&\frac{\d S(\ck)}{\d b}-S_{\ck}\left(\frac{\d\ck}{\d b}-\pa^\m A_\m\right)=
\cg(\ck)~,~~
\frac{\d S(\ck)}{\d\p^0}-S_{\ck}\left(\frac{\d\ck}{\d \p^0}-
\pa^\m B_\m\right)=\cg^1(\ck)~,\nonumber\\
&&\frac{\d S(\ck)}{\d\p^{0\n}}-S_{\ck}\left(\frac{\d\ck}{\d \p^{0\n}}-
\pa^\m K_{\m\n}-\pa_\n\r^0\right)=\cg^1_\n(\ck)+\pa_\n\l^1~,~~
\frac{\d S(\ck)}{\d \p^{-1}}+S_{\ck}\left(\frac{\d\ck}{\d \p^{-1}}-
\pa^\m B^1_\m-\l^1\right)=\cg^2(\ck)~,\nonumber\\
&&\cg S(\ck)+S_{\ck}\cg(\ck)=0~,~~
\cg^1 S(\ck)+S_{\ck}\cg^1(\ck)=0~,\nonumber\\
&&\cg^1_\m S(\ck)+S_{\ck}(\cg^1_\m(\ck)+\pa_\m\l^1)=0~,~~
\cg^2 S(\ck)-S_{\ck}\cg^2(\ck)=0~,\nonumber\\
&&\frac{\d S(\ck)}{\d \r^0}-S_{\ck}\left(\frac{\d\ck}{\d \r^0}+
\pa^\m \p^0_\m\right)=0~,~~
\frac{\d S(\ck)}{\d \l^1}+S_{\ck}\left(\frac{\d\ck}{\d \l^1}+
\pa^\m \bc^{-1}_\m+\p^{-1}\right)=\frac{\d\ck}{\d \r^0}+
\pa^\m \p^0_\m~,\nonumber\\
&&\ov\cg S(\ck)+S_{\ck}(\ov\cg(\ck)-\D^{\ov\cg})=\ov\cf(\ck)~,~~
\ov\cf S(\ck)-S_{\ck}\ov\cf(\ck)=0~,\nonumber\\
&&\AE S(\ck)-S_{\ck}(\AE(\ck)-\D^\AE)=\O(\ck)-\D^\O~,~~
\O S(\ck)+S_{\ck}(\O(\ck)-\D^\O)=0~;
\label{algebra1}
\ea
\item Other identities
\end{enumerate}
\ba
&&\{\cw_\m,\cw_\n\}=0~,~~\{\cw_\m,{\L}\}=0~,~~\{{\L},{\L}\}=0~,\nonumber\\
&&\cw_{\m}(\ov\cg(\ck)-\D^{\ov\cg})+\ov\cg(\cw_{\m}(\ck)-\D^{\cw_\m}_{\m})=0~,~~
\L(\ov\cg(\ck)-\D^{\ov\cg})+\ov\cg({\L}(\ck)-\D^{\L})=\AE(\ck)-\D^\AE~,
\nonumber\\
&&\cw_\m\ov\cf(\ck)-\ov\cf(\cw_\m(\ck)-\D^{\cw_\m}_\m)=0~,~~
\L\ov\cf(\ck)-\ov\cf(\L(\ck)-\D^{\L})=\O(\ck)-\D^\O~,\nonumber\\
&&\cw_{\m}(\AE(\ck)-\D^\AE)-\AE(\cw_{\m}(\ck)-\D^{\cw_\m}_{\m})=0~,~~
\L(\AE(\ck)-\D^\AE)-\AE(\L(\ck)-\D^{\L})=0~,\nonumber\\
&&\cw_{\m}(\O(\ck)-\D^\O)+\O(\cw_{\m}(\ck)-\D^{\cw_\m}_{\m})=0~,~~
\L(\O(\ck)-\D^\O)+\O(\L(\ck)-\D^{\L})=0~,\nonumber\\
&&\ov\cg^a(\ov\cg^b(\ck)-\D^{\ov\cg b})+\ov\cg^b(\ov\cg^a(\ck)
-\D^{\ov\cg a})=0~,~~
\ov\cf^a\ov\cf^b(\ck)-\ov\cf^b\ov\cf^a(\ck)=0~,~~
\ov\cg^a\ov\cf^b(\ck)-\ov\cf^a(\ov\cg^b(\ck)-\D^{\ov\cg b})=0~,\nonumber\\
&&\ov\cg^a(\AE^b(\ck)-\D^{\AE b})-\AE^a(\ov\cg^b(\ck)-\D^{\ov\cg b})=0~,~~
\ov\cg^a(\O^b(\ck)-\D^{\O b})+\O^a(\ov\cg^b(\ck)-\D^{\ov\cg b})=0~,\nonumber\\
&&\ov\cf^a(\AE^b(\ck)-\D^{\AE b})-\AE^a\ov\cf^b(\ck)=0~,~~
\ov\cf^a(\O^b(\ck)-\D^{\O b})-\O^a\ov\cf^b(\ck)=0~,\nonumber\\
&&\AE^a(\AE^b(\ck)-\D^{\AE b})-\AE^b(\AE^a(\ck)-\D^{\AE a})=0~,~~
\O^a(\O^b(\ck)-\D^{\O b})+\O^b(\O^a(\ck)-\D^{\O a})=0~,\nonumber\\
&&\AE^a(\O^b(\ck)-\D^{\O b})-\O^a(\AE^b(\ck)-\D^{\AE b})=0~,\nonumber\\
&&[\cp_\m,\Q]=0~\forall\Q\in\{S_{\ck},\cw_\m,\L,\cg,\cg^1,\cg^1_\m,\cg^2,
\ov\cg,\ov\cf,\AE,\O,\cp_\m\}~,
\label{algebra2}
\ea
where $\cp_\m$ is the Ward operator associated to translations:
\be
\cp_\m=\sum_\vf{\mbox{Tr}}\int d^3x~\pa_\m\vf\frac{\d}{\d \vf}~.
\ee
The first group of identities involving the Slavnov-Taylor operator 
given by (\ref{algebra1}) are those which yield the conditions (the 
well-known Wess-Zumino consistency condition is one of them) to be 
satisfied by the quantum breaking of the Slavnov-Taylor identity 
(\ref{slavnov}) allowed by the Quantum Action Principle \cite{piguet1}.

\section{Finiteness}
In this Section is summarized the results on the study of stability of 
the classical action under radiative corrections, the implementation of 
the Slavnov-Taylor identity at the quantum level and the conclusion on the
finiteness of the BFK-model at all orders in perturbation theory.    

\subsection{Stability}\label{stability}
In order to check whether the action in the tree-approximation is 
stable under radiative corrections or not, we perturb it by an
arbitrary integrated local functional $\S^c$, such that
\be
\wt\G^{(0)}=\G^{(0)}+\ve \S^c~, \label{adef}
\ee
where $\ve$ is an infinitesimal parameter. The functional $\S^c$ has the 
same quantum numbers\footnote{See TABLE \ref{dimensions} for the dimension 
and the ghost number of all fields and anti-fields.} 
(dimension $3$, Faddeev-Popov charge $0$, \dots) as the classical action, 
$\G^{(0)}$, and the deformed action, $\wt\G^{(0)}$, must obey in the same way 
all the identities $\G^{(0)}$ does, leading therefore, to the following 
homogeneous conditions to be fulfilled by the counterterm, $\S^c$: 
\ba
&&\cs_{\G^{(0)}}\S^c=\cw_\m\S^c=\L\S^c=\frac{\d\S^c}{\d b}
=\frac{\d\S^c}{\d \p^0}=\frac{\d\S^c}{\d \p^{0\m}}
=\frac{\d\S^c}{\d \p^{-1}}=\frac{\d\S^c}{\d \r^0}=\frac{\d\S^c}{\d \l^1}=0
~, \nonumber \\
&&\cg\S^c=\cg^1\S^c=\cg^1_\m\S^c=\cg^2\S^c=\ov\cg\S^c=\AE\S^c=\ov\cf\S^c
=\O\S^c=0~.
\ea

In searching for the most general counterterm satisfying all the constraints 
listed above, it can be shown \cite{paper} that there is no integrated local 
polynomial in the fields which survives those requirements, then
\be
\S^c=0~,
\ee 
meaning that the usual ambiguities due to the renormalization procedure do 
not appear in the BFK-model. It remains now to prove the absence of anomalies 
in order to conclude about the perturbative finiteness of the model. 

\subsection{Anomaly and finiteness}\label{anomaly}
At the quantum level the vertex functional, $\G$, which coincides with 
the classical action (\ref{bfkaction}), $\G^{(0)}$, at order 0 in $\hbar$,
\be
\G=\G^{(0)}+{\co}(\hbar)~,\label{vertex}
\ee
has to satisfy the same constraints as the classical action does. However, 
according to the Quantum Action Principle \cite{piguet1} the 
Slavnov-Taylor identity (\ref{slavnov}) gets a quantum breaking
\be
\cs(\G)=\D\cdot\G=\D+{\co}(\hbar \D)~, 
\label{slavnovbreak}
\ee
where $\D$ is an integrated local functional with ghost number 1 and 
dimension $3$. The absence of anomalies amounts to show that the 
Slavnov-Taylor identity can be implemented at the quantum level 
at the expenses of a BRS-trivial breaking $\D=\cs_{\G}{\wh\D}^{(0)}$, 
called noninvariant counterterm.

The nilpotency identity, $S_{\G}S(\G)=0$, together with
\be
\cs_{\G}=\cs_{\G^{(0)}}+{\co}(\hbar)~,
\ee
implies the Wess-Zumino consistency condition for the breaking $\D$:
\be
\cs_{\G^{(0)}}\D=0~,\label{breakcond1}
\ee
beyond that, through the algebra (\ref{algebra1}), $\D$ satisfies:
\ba
&&\cw_\m\D=\L\D=\frac{\d\D}{\d b}
=\frac{\d\D}{\d \p^0}=\frac{\d\D}{\d \p^{0\m}}
=\frac{\d\D}{\d \p^{-1}}=\frac{\d\D}{\d \r^0}=\frac{\d\D}{\d \l^1}=0
~, \nonumber \\
&&\cg\D=\cg^1\D=\cg^1_\m\D=\cg^2\D=\ov\cg\D=\AE\D=\ov\cf\D=\O\D=0~.
\label{breakcond2}
\ea

The Wess-Zumino consistency condition (\ref{breakcond1}) constitutes a 
cohomology problem in the sector of ghost number one. 
Its solution can always be written as a sum of a trivial cocycle 
$\cs_{\G^{(0)}}{\wh\D}^{(0)}$, where ${\wh\D}^{(0)}$ has ghost number $0$, 
and of nontrivial elements belonging to the cohomology of $\cs_{\G^{(0)}}$ 
(\ref{slavnovlin}) in the sector of ghost number one, $\ca^{(1)}$:
\be
\D=\ca^{(1)}+\cs_{\G^{(0)}}{\wh\D}^{(0)}~. 
\label{breaksplit}
\ee
Although the constraints imposed to $\D$ in (\ref{breakcond1}) and 
(\ref{breakcond2}) show that $\ca^{(1)}=0$ \cite{paper}, it can be 
proved quite generally that in three-dimensions there is no anomaly, 
since the cohomology in the sector of ghost 
number $1$ is empty up to possible terms in the Abelian ghosts 
\cite{kreuzer}. However, through the arguments of \cite{bbbc} 
we conclude that the $U(1)$-ghosts do not contribute to the anomaly due 
to their freedom or soft coupling, then the Slavnov-Taylor identity 
is implemented at the quantum level.  

In conclusion, the absence of counterterms in the study of stability, as 
presented in Subsection \ref{stability}, together with the result 
of Subsection \ref{anomaly} concerning the absence of anomaly lead to 
a proof on the finiteness of the BFK-model at all orders in perturbation 
theory.

\begin{table}[t]
\begin{center}
\begin{tabular}{|c||c|c|c|c|c|c|c|c|c|c|c|c|c|}
\hline
& $B_\m$ & $K_{\m\n}$ & $A_\m$ & $\f$ & $c$ & $B^2$ & $B^1$ & $B^1_\m$ & $\bc$ 
& $b$ & $\bc^{-1}$ & $\p^0$ & $\bc^{-1}_\m$\\ 
\hline\hline
$d$ & $1$ & $1$ & $1$ & $1$ & $0$ & $-1$ & $0$ & $0$ & $1$ & $1$ & $1$ & $1$ &
$1$ \\ \hline
$\Phi\Pi$ & $0$ & $0$ & $0$ & $0$ & $1$ & $2$ & $1$ & $1$ & $-1$ & $0$ & $-1$ &
$0$ & $-1$\\ 
\hline
\hline
& $\p^0_\m$ & $\bc^{-2}$ & $\p^{-1}$ & $\r^0$ & $\l^1$ & $B^*_\m$ & 
$K^*_{\m\n}$ & $A^*_\m$ & $\f^*$ & $c^*$ & $B^{2*}$ & $B^{1*}$ & $B^{1*}_\m$\\ 
\hline\hline
$d$ & $1$ & $2$ & $2$ & $1$ & $1$ & $2$ & $2$ & $2$ & $2$ & $3$ & $4$ & $3$ & 
$3$ \\ \hline
$\Phi\Pi$ & $0$ & $-2$ & $-1$ & $0$ & $1$ & $-1$ & $-1$ & $-1$ & $-1$ & $-2$ &
$-3$ & $-2$ & $-2$\\  
\hline
\end{tabular}
\end{center}
\caption[]{\label{dimensions}Dimension $d$ and ghost number $\F\P$.}
\end{table}

\underline{Acknowledgements}: 
One of the authors (O.M.D.C.) thanks Jos\'e A. Helay\"el-Neto, Susana I.Z. 
Caride, the Head of DCP-CBPF and Anibal O. Caride, the Head of CFC-CBPF, 
for financial support and the warm hospitality during all his visits at 
the {\it Departamento de Teoria de Campos e Part\'\i culas (DCP)} of the 
{\it Centro Brasileiro de Pesquisas F\'\i sicas (CBPF)}. He
dedicates this work to his wife, Zilda Cristina, to his kids, Vittoria 
and Enzo, and his mother, Victoria. (J.M.G.) thanks Harald Ita for many 
useful discussions.  

\end{document}